\documentclass[11pt]{article}
\pdfoutput=1
\usepackage[sumlimits,intlimits,namelimits]{amsmath}
\usepackage{amssymb}
\usepackage[english]{babel}
\usepackage{bbm}
\usepackage{parskip}

\usepackage[margin=1cm]{caption}
\usepackage{enumitem}

\usepackage{graphicx}

\usepackage{cite}

\usepackage[makeroom]{cancel}
\usepackage[usenames,dvipsnames]{color}
\usepackage[dvipsnames]{xcolor}

\usepackage[bordercolor=black,backgroundcolor=yellow,linecolor=black,
textsize=scriptsize,shadow]{todonotes}

\global\parskip 6pt

\graphicspath{{pics/}}

\usepackage{hyperref}
\hypersetup{
    colorlinks=true,
    linkcolor=blue,
    filecolor=magenta,      
    urlcolor=cyan,
}

\renewcommand{\thefootnote}{\fnsymbol{footnote}}

\begin{document}

\begin{titlepage}
\vspace*{.5cm}
\begin{center}
{\Large{\bf On the efficacy of virtual seminars}
} \\[.5ex]
\vspace{1cm}
\emph{Gravity, Quantum Fields and Information} research group\footnote{Web: \href{http://aei.mpg.de/GQFI}{aei.mpg.de/GQFI}; Contact: \href{mailto:GQFI@aei.mpg.de}{GQFI@aei.mpg.de}; Members: Hugo Camargo, Michal P. Heller (PI), Ro Jefferson, Johannes Knaute, Ignacio Reyes, Sukhbinder Singh, and Viktor Svensson.}\vspace{0.5cm}\\
\em
Max Planck Institute for Gravitational Physics (Albert Einstein Institute)\\
Am M\"uhlenberg 1, D-14476 Potsdam-Golm, Germany
\end{center}
\vspace*{1cm}
\begin{abstract}
    During the SARS-CoV-2 pandemic, theoretical high-energy physics, and likely also the majority of other disciplines, are seeing a surge of virtual seminars as a primary means for scientific exchange. In this brief article, we highlight some compelling benefits of virtualizing research talks, and argue for why virtual seminars should continue even after the pandemic. Based on our extensive experience on running online talks, we also summarize some basic guidelines on organizing virtual seminars, and suggest some directions in which they could evolve.
\end{abstract}
\vspace*{.25cm}
\end{titlepage}


\setcounter{tocdepth}{2}
\setcounter{secnumdepth}{4}
\clearpage

\section{Introduction and overview}

\setcounter{footnote}{0} 
\renewcommand{\thefootnote}{\arabic{footnote}}

One of the main driving forces behind scientific progress is the free and timely exchange of ideas. A real breakthrough in this direction happened already about 30 years ago with the launch of the \emph{arXiv} preprint server, which allowed for a free, timely, and scalable dissemination of the newest knowledge presented in research papers. Despite the enormous progress in technology since that time however, other equally essential channels of research dissemination – namely, research seminars and conferences – do not seem to have evolved proportionally to technological advancement.

The advent of the SARS-CoV-2 pandemic has, however, dramatically laid bare the severe limitations of traditional seminars and conferences\footnote{We are writing this article from the perspective of our field, i.e. theoretical physics at the intersection of gravity, quantum field theory, and quantum information science. The relevant \emph{arXiv} category is primarily \emph{hep-th} and to a lesser extent \emph{quant-ph}, \emph{nucl-th} and \emph{gr-qc}. Also, the scientific environment we refer to is Germany, where our group is located.\\}.

As a result of SARS-CoV-2, several scientific meetings and visits have since been canceled and replaced by a tangible sense of vacuum and lack of established alternatives. But innovative solutions have emerged. To keep science progressing in these difficult times, an increasing number of groups are swiftly migrating to online communication platforms. Several groups have, in fact, already upgraded their regular seminars to virtual talks, and the first virtual conferences have been organized, with more brewing in the pipeline. But one wonders if groups and communities will resort to their former habits after the pandemic has passed, or whether virtual seminars and conferences will permanently join our traditional array of science communication channels.

In this brief article, we outline reasons for why we believe that virtual seminars and conferences might, or rather should, be here to stay.

Ours was one of the first groups in our discipline to have launched an entirely virtual seminar series long before the current pandemic appeared on the horizon. In the past 30 months, we have hosted more than 80 virtual talks. Since the first broadcast of our group seminar, we have also launched three new specialized seminar series in collaboration with partner groups: 
\begin{enumerate}[label=(\roman*)]
\item \emph{Virtual Seminars on Quantum Gravity and Information} (\href{http://QGIseminar.org}{QGIseminar.org}) in partnership with \emph{the Amsterdam String Theory group},
\item \emph{Virtual Seminar on Tensor Networks in High Energy Physics} (\href{http://hepTNseminar.org}{hepTNseminar.org}) in collaboration with \emph{DESY Zeuthen}, \emph{MPQ}, and \emph{Perimeter Institute},
\item \emph{Virtual Seminar on t-, T- and $\mu$-dependence in Quantum Field Theory} (\href{http://tTmuQFTseminar.org}{tTmuQFTseminar.org}).
\end{enumerate}
The recording of the vast majority of our virtual talks are made available afterward on our \emph{YouTube} channel (\href{https://www.youtube.com/c/GravityQuantumFieldsandInformationAEI}{youtube.com/c/GravityQuantumFieldsandInformationAEI}) and are free to watch by anyone. Some of us are also presently organizing a virtual conference on \emph{Complexity from Quantum Information to Black Holes}\footnote{\href{https://www.lorentzcenter.nl/complexity-from-quantum-information-to-black-holes.html}{www.lorentzcenter.nl/complexity-from-quantum-information-to-black-holes.html}} under the aegis of Leiden’s \emph{Lorentz Centre} in~June~2020.

The entire machinery and impact of virtual communications and how they relate to traditional methods of communication is, of course, an active area of academic research in social sciences. Our views presented here are based entirely on the first hand experience that we have accumulated from our virtual seminars. The purpose of this article is to make use of this timely opportunity to share a vision for science communication far into the future, beyond the current circumstances. It seems clear to us that the current situation has brought forth an important message, namely, virtual seminars and conferences are not only an alternative mechanism for science communication, but significantly broaden its scope and capacity.

Before discussing the benefits of virtual seminars and conferences, it is perhaps necessary to highlight some common myths behind why the academic community has largely been less enthused by virtual communication. (This is not to say that there are no glitches in the virtual format. We mention some later.) First, it is often imagined that virtual talks must be of lower quality and impact when compared to real seminars. However, from our experience, this claim seems exaggerated. The relevant online communication tools have been around for several years now, and they offer a remarkably smooth and highly optimized interface for virtual talks. The virtual format also seems to be equally effective for black-board (or digital board) lectures as it is for power-point presentations, thus contradicting another common myth. Another common criticism is that it is harder to absorb information when watching talks online, based on noted differences between how we consume information from personal interactions versus online content. But it is useful to remind here that online content is available for repeated viewing.

Below we first discuss some obvious as well as some perhaps not so apparent benefits for science communication. We then discuss the manifest and readily quantifiable contribution that virtual seminars and conferences can make to climate change. We conclude with some useful guidelines that we have learned from organizing virtual seminars in the last two years.

\section{Benefits for science communication}

\begin{enumerate}
    \item \textbf{Reduced cost:} Perhaps the most apparent benefit is saving taxpayers’ money. If we assume that organizing an average speaker’s visit within Europe costs 500 EUR, a rather modest estimate, then eighty virtual talks save costs amounting to more than one year of a Ph.D. student salary in Germany, or equally a significant fraction of a post-doc salary. One can therefore imagine that three geographically- and scientifically-close group virtualize their talks and invest saved resources into hiring a joint Ph.D. student or a post-doc. The savings from curbing long-haul travel are, of course, much more pronounced. (Our virtual seminars have had fifty speakers from the Americas and ten from East Asia.) Streaming a seminar also obviates the need to present the same talk multiple times to geographically disconnected audiences, and thus there are multiplicative cost savings for every talk. Such savings can be crucial for small-scale groups with fixed, restricted budgets such as ours. They can also provide additional employment opportunities in a field where positions are scarce and the number of applicants large. We also note that the costs of organizing traditional conferences can be extremely high.
    \item \textbf{Ease of organization and timely research dissemination:} Virtual seminars and conferences can also save the sometimes massive time and effort devoted to organizing academic visits. The reduced organization time not only eases administrative workload but also demonstrably leads to timely dissemination of new research. Many speakers of our virtual seminars were able to present a paper within a couple weeks of releasing it on \emph{arXiv}. In contrast, there is usually a substantial time gap between the release of new research and presenting it in-person at a seminar or conference. Thus, virtual seminars can help disseminate new research faster.
    \item \textbf{Scale and quality of presentation:} The next obvious benefit is that virtual seminars can reach out to a much larger global audience across different time zones. For example, one of our recent \emph{QGI seminars} had 180 participants, of the order of a medium-sized conference in our discipline. An easily accessible large-sized audience can accelerate the visibility of new research and enhance scientific networking. Moreover, virtual mediums could also enable broadcasting colloquium style talks to the general public, which can help scale-up scientific outreach more generally. (In fact, \emph{Perimeter Institute} has already been broadcasting its on-site outreach talks for quite some time.) Access to scientific seminars can also be of immeasurable benefit to less privileged research audiences from developing or geographically isolated countries.
    
    We have already noted that, somewhat unexpectedly, the quality of a virtual seminar does not seem to be compromised in any meaningful sense when compared to in-person delivery. In fact, there seem to be some small but significant advantages. Firstly, the visual and audio delivery of a virtual seminar can be more effective since it offers everyone in the audience the “first-row” experience. (In a popular seminar or a large conference, latecomers often find themselves straining to catch the presentation at the end of a large room.) It is also simpler to use props in a virtual seminar: one can simply deliver the talk from where the accessories are located instead of carrying them to the audience. Imagine an experimentalist giving a talk directly from their lab or from the LHC, or a theoretical physicist showcasing a real-time simulation on a supercomputer or on their local computing cluster.  This may seem like belaboring a point, but it illustrates very clearly that virtual seminars are not merely replacements for traditional seminars, but can potentially serve as a platform for a more vibrant and impactful delivery.
    \item \textbf{Towards a global online video \emph{arXiv}:} Virtual seminars are much more quickly recorded than in-person seminars, and do not require expensive camera and recording equipment. Some available software solutions allow recording seminars locally on any participating computer with the press of a button. If virtual seminars take-off as a regular scientific activity, it is not hard to envision the emergence of a shared catalog, similar to \emph{arXiv} for research papers, of links to past and upcoming virtual seminars across the globe. Links could be posted by research groups anywhere in the world. (Some informal catalogs, e.g., social media pages, are beginning to appear, but they are not conveniently organized for scalable access.) At some stage, it would also be natural to incorporate individual submissions, that is, a single researcher can  record a presentation of their work and submit the link. (This may require moderation of some sort.) One new possibility on this front is providing either a modular format for a talk of, e.g., 15 minutes of introduction and summary and, e.g., 30 minutes of detailed explanations, or even recording one very short talk with the key punchline (an analogue of a letter) and a longer talk explaining details (an analogue of a technical paper).
    
    Several physics institutes, such as \emph{Perimeter Institute} in Canada and \emph{Kavli Institute for Theoretical Physics} in USA, already maintain substantial video repositories of their local seminars\footnote{Repectively: \href{http://pirsa.org}{pirsa.org} and \href{https://www.kitp.ucsb.edu/online-talks}{www.kitp.ucsb.edu/online-talks}.}, which have proved to be a tremendously valuable resource. But these recordings are of seminars that were presented by researchers either working at or visiting these institutes. Therefore, such repositories are less encompassing and can only grow at a rate that is once again limited by the frequency of academic visits; thus, they do not currently serve to disseminate new research in such a timely and unrestricted way.
\end{enumerate}

\section{Contribution to climate change activism}

Perhaps the most significant environmental impact of theoretical physics comes from academic travel. Large-scale virtual talks can dramatically reduce the associated carbon footprint. Our own virtual seminar series provides a small-scale but encouraging illustration. If we had instead opted to organize all the seminars as traditional on-site visits, we would have had to bring twenty-four speakers from Europe, fifty from the Americas, and ten from Asia. A crude estimate of the carbon footprint from flights would correspond to seventy tons of carbon dioxide in under two and a half years. This emission, according to publicly available data\footnote{See, e.g., \href{https://data.worldbank.org/indicator/en.atm.co2e.pc}{data.worldbank.org/indicator/en.atm.co2e.pc}.}, equates to around thirty tons of carbon dioxide per year, which is roughly speaking twice the contribution of an average North American and three times more than a European. The above estimates ignore the emission costs of the global computing and transmission infrastructure, which are undoubtedly significant (back-of-the envelope calculations estimate them to be of the order of 10\% of the saved emissions), but more easily transitioned to green energy than air travel.

\section{Guidelines for organizing a virtual seminar}

The experience of giving and listening to a virtual seminar is, of course, different from that of a traditional seminar. It may be a particularly odd experience for some speakers, given the absence of visual cues about how the seminar is progressing. Below we share some guidelines that we learned while developing our virtual seminars, which might be useful for groups who are contemplating transitioning to virtual seminars.

\subsection{Set-up}
\begin{enumerate}[label=\alph*)]
\item Our seminars were broadcast only to active researchers who subscribed to our mailing list.
\item The virtual seminar room link was only shared with subscribed participants by email, and was not listed on our seminar advertisements on public domains.
\item Seminars were advertised at least a week in advance, and a reminder was sent on the day of the seminar.
\item We invited the speakers to set-up their talk at least 15-30 minutes before the talk.
\end{enumerate}

\subsection{Format of the talks}
\begin{enumerate}[label=\alph*)]
\item Seminar lengths of up to 1 hour (including questions) seemed to work well. Most people were usually happy to stay up to an hour in our seminars, but longer talks did not work as well. We also experimented with 45-minute seminars and found that they were significantly more comfortable to watch and follow.
\item Digital board talks generally worked better than blackboard talks, due to better lighting and readability.
\item Blackboard talks work better if the speaker divides the blackboard into sections and manually adjusts the camera position to zoom to one section at a time. (The entire board is difficult to read on a computer screen.)
\end{enumerate}

\subsection{Delivery}
\begin{enumerate}[label=\alph*)]
\item We asked speakers to pause periodically in their presentations to break for any questions explicitly. Sometimes including "blank" slides to break for questions at regular intervals helped regulate the pace of the talk.
\item To improve the lack of visual cues, both for the speaker and the audience, we felt that it was better to include a live visual (in a small window at the top corner) of the speaker while they were speaking, and also of some of the organizers (for the speaker to collect cues).
\end{enumerate}

\subsection{Moderation}
\begin{enumerate}[label=\alph*)]
\item We found it better if people virtually raise hands (e.g., using such a built-in facility in the relevant software) before a moderator calls out their name to ask their question.
\item It is better to make any announcements at the beginning of the talk, as the audience tends to dissipate very quickly at the end of a virtual seminar.
\item We announced at the start of the seminar that the seminar will be recorded and uploaded on our \emph{YouTube} channel.
\item Excessive text exchanges over publicly available during the seminar chat are distracting.
\end{enumerate}

\subsection{Technological glitches}
\begin{enumerate}[label=\alph*)]
\item If the software allows, setting up multiple online hosts for the seminar helps in case there is a technological failure at any one end.
\end{enumerate}

\section{Closing remarks}
In the end, we would like to stress that we do not, of course, believe that all scientific meetings must be virtual. In particular, the benefits of the human interaction component of traditional seminars and conferences can hardly be neglected. But we do suggest to keep using the virtual medium of scientific exchange whenever deep interactions are not necessary, and to maximize the impact (and cost-benefit) of in-person seminars by sharing them virtually.

\vspace{10 pt}

\noindent\textbf{Acknowledgements:} Our group is supported by \emph{the Alexander von Humboldt Foundation} and \emph{the Federal Ministry for Education and Research} through \emph{the Sofja Kovalevskaja Award}. Our vision for virtual seminars would have never materialized if not help from AEI's excellent and dedicated IT Team, to which we are very grateful.

\end{document}